\documentstyle[emulateapj,rotate,times]{article}
 
\newcommand{\etal}{et~al.\ }
\newcommand{\PVdblt}{{\rm P}\kern 0.1em{\sc v}~$\lambda\lambda 1117, 1128$}
\newcommand{\CaIIdblt}{{\rm Ca}\kern 0.1em{\sc ii}~$\lambda\lambda 3934, 3969$}
\newcommand{\AlIIIdblt}{{\rm Al}\kern 0.1em{\sc iii}~$\lambda\lambda 1854, 1862$}
\newcommand{\CIVdblt}{{\rm C}\kern 0.1em{\sc iv}~$\lambda\lambda 1548, 1550$}
\newcommand{\MgIIdblt}{{\rm Mg}\kern 0.1em{\sc ii}~$\lambda\lambda 2796, 2803$}
\newcommand{\NVdblt}{{\rm N}\kern 0.1em{\sc v}~$\lambda\lambda 1238, 1242$}  
\newcommand{\SVIdblt}{{\rm S}\kern 0.1em{\sc vi}~$\lambda\lambda 933, 944$} 
\newcommand{\OVIdblt}{{\rm O}\kern 0.1em{\sc vi}~$\lambda\lambda 1031, 1037$} 
\newcommand{\SiIIdblt}{{\rm Si}\kern 0.1em{\sc ii}~$\lambda\lambda 1190, 1193$} 
\newcommand{\SiIVdblt}{{\rm Si}\kern 0.1em{\sc iv}~$\lambda\lambda 1393, 1402$} 
\newcommand{\PV}{\hbox{{\rm P}\kern 0.1em{\sc v}}}
\newcommand{\AlI}{\hbox{{\rm Al}\kern 0.1em{\sc i}}}
\newcommand{\AlII}{\hbox{{\rm Al}\kern 0.1em{\sc ii}}}
\newcommand{\AlIII}{{\hbox{\rm Al}\kern 0.1em{\sc iii}}}
\newcommand{\CaII}{\hbox{{\rm Ca}\kern 0.1em{\sc ii}}}
\newcommand{\CII}{\hbox{{\rm C}\kern 0.1em{\sc ii}}}
\newcommand{\CIIe}{\hbox{{\rm C$^{\ast}$}\kern 0.1em{\sc ii}}}
\newcommand{\CIII}{\hbox{{\rm C}\kern 0.1em{\sc iii}}}
\newcommand{\CIV}{\hbox{{\rm C}\kern 0.1em{\sc iv}}}
\newcommand{\CV}{\hbox{{\rm C}\kern 0.1em{\sc v}}}
\newcommand{\HI}{\hbox{{\rm H}\kern 0.1em{\sc i}}}
\newcommand{\HII}{\hbox{{\rm H}\kern 0.1em{\sc ii}}}
\newcommand{\Lya}{\hbox{{\rm Ly}\kern 0.1em$\alpha$}}
\newcommand{\Lyb}{\hbox{{\rm Ly}\kern 0.1em$\beta$}}
\newcommand{\Lyg}{\hbox{{\rm Ly}\kern 0.1em$\gamma$}}
\newcommand{\Lyd}{\hbox{{\rm Ly}\kern 0.1em$\delta$}}
\newcommand{\Lye}{\hbox{{\rm Ly}\kern 0.1em$\epsilon$}}
\newcommand{\Lyphi}{\hbox{{\rm Ly}\kern 0.1em$\phi$}}
\newcommand{\Lyfive}{\hbox{{\rm Ly}\kern 0.1em$5$}}
\newcommand{\Lysix}{\hbox{{\rm Ly}\kern 0.1em$6$}}
\newcommand{\Lyseven}{\hbox{{\rm Ly}\kern 0.1em$7$}}
\newcommand{\Lyeight}{\hbox{{\rm Ly}\kern 0.1em$8$}}
\newcommand{\Lynine}{\hbox{{\rm Ly}\kern 0.1em$9$}}
\newcommand{\Lyten}{\hbox{{\rm Ly}\kern 0.1em$10$}}
\newcommand{\Lyeleven}{\hbox{{\rm Ly}\kern 0.1em$11$}}
\newcommand{\HeI}{\hbox{{\rm He}\kern 0.1em{\sc i}}}
\newcommand{\HeII}{\hbox{{\rm He}\kern 0.1em{\sc ii}}}
\newcommand{\FeI}{\hbox{{\rm Fe}\kern 0.1em{\sc i}}}
\newcommand{\FeII}{\hbox{{\rm Fe}\kern 0.1em{\sc ii}}}
\newcommand{\FeIII}{\hbox{{\rm Fe}\kern 0.1em{\sc iii}}}
\newcommand{\MnII}{\hbox{{\rm Mn}\kern 0.1em{\sc ii}}}
\newcommand{\MgI}{\hbox{{\rm Mg}\kern 0.1em{\sc i}}}
\newcommand{\MgII}{\hbox{{\rm Mg}\kern 0.1em{\sc ii}}}
\newcommand{\MgIII}{\hbox{{\rm Mg}\kern 0.1em{\sc iii}}}
\newcommand{\NI}{\hbox{{\rm N}\kern 0.1em{\sc i}}}
\newcommand{\NII}{\hbox{{\rm N}\kern 0.1em{\sc ii}}}
\newcommand{\NIII}{\hbox{{\rm N}\kern 0.1em{\sc iii}}}
\newcommand{\NV}{\hbox{{\rm N}\kern 0.1em{\sc v}}}
\newcommand{\OVI}{\hbox{{\rm O}\kern 0.1em{\sc vi}}}
\newcommand{\OI}{\hbox{{\rm O}\kern 0.1em{\sc i}}}
\newcommand{\OII}{\hbox{[{\rm O}\kern 0.1em{\sc ii}]}}
\newcommand{\OIV}{\hbox{{\rm O}\kern 0.1em{\sc iv}]}}
\newcommand{\SI}{{\rm S}\kern 0.1em{\sc i}}
\newcommand{\SIV}{{\rm S}\kern 0.1em{\sc iv}}
\newcommand{\SVI}{{\rm S}\kern 0.1em{\sc vi}}
\newcommand{\SiI}{\hbox{{\rm Si}\kern 0.1em{\sc i}}}
\newcommand{\SiII}{\hbox{{\rm Si}\kern 0.1em{\sc ii}}}
\newcommand{\SiIII}{\hbox{{\rm Si}\kern 0.1em{\sc iii}}}
\newcommand{\SiIV}{\hbox{{\rm Si}\kern 0.1em{\sc iv}}}
\newcommand{\SII}{\hbox{{\rm S}\kern 0.1em{\sc ii}}}
\newcommand{\SIII}{\hbox{{\rm S}\kern 0.1em{\sc iii}}}
\newcommand{\NaI}{\hbox{{\rm Na}\kern 0.1em{\sc i}}}
\newcommand{\TiII}{\hbox{{\rm Ti}\kern 0.1em{\sc ii}}}
\newcommand{\kms}{\hbox{km~s$^{-1}$}}
\newcommand{\cmsq}{\hbox{cm$^{-2}$}}

\tighten
\begin{document}
 
 
\lefthead{CHARLTON ET~AL.}
\righthead{CONSTRAINTS ON INTRAGROUP HVCs}


\title{QSO Absorption Line Constraints on Intragroup High--Velocity
Clouds}

\author{Jane~C.~Charlton\altaffilmark{1}, Christopher~W.~Churchill,
and Jane~R.~Rigby}
\affil{The Pennsylvania State University, University Park, PA 16802 \\
charlton, cwc, jrigby@astro.psu.edu}

\altaffiltext{1}{Center for Gravitational Physics and Geometry}

\begin{abstract}
We show that the number statistics of moderate redshift {\MgII} and
Lyman limit absorbers may rule out the hypothesis that high velocity
clouds are infalling intragroup material.
\end{abstract}



\section{Introduction}

The origin(s) of high--velocity clouds (HVCs), gaseous material that
departs from the Galactic rotation law by more than 100~{\kms}, is a
topic under debate.
Undoubtedly, some HVCs arise from tidal streams (e.g.\ the Magellanic
Stream), and from fountain processes local to the Galaxy
(\cite{wakker97}).  
Recently, however, the hypothesis that {\it most\/} HVCs are
distributed ubiquitously throughout the Local Group and are relics of
group formation has returned to favor (\cite{blitz};
\cite{braun-burton}).

In the intragroup HVC hypothesis
(1) the cloud kinematics follow the Local Group standard of
rest (LGSR), not the Galactic standard of rest (GSR), with
the exception of some HVCs related to tidal stripping or
Galactic fountains (\cite{blitz}; \cite{braun-burton}); 
(2) the cloud Galactocentric distances are typically 1~Mpc; 
(3) the extended HVC cloud complexes are presently
accreting onto the Milky Way;
(4) the clouds are local analogs of the Lyman limit absorbers observed in
quasar spectra; 
(5) the clouds have masses of $10^{7}$~M$_{\odot}$ and greater; and 
(6) the metallicities are lower than expected if the material
originated from blowout or fountains from the Milky Way
(\cite{wakker99}; \cite{bowen93}).

Blitz \etal (1999\nocite{blitz}; hereafter BSTHB) suggest that there
are $\sim 300$ clouds above the $21$--cm $N({\HI})$ detection threshold of 
$\simeq 2 \times 10^{18}$~{\cmsq}.   These clouds have radii $\sim 15$~kpc 
and are ubiquitous throughout the group.  

Braun and Burton (1999\nocite{braun-burton}, hereafter BB) 
cataloged $65$ Local Group CHVCs, which represent a 
homogeneous subset of the HVC population discussed by BSTHB.
High resolution $21$--cm observations (\cite{bb-hires})
show that the CHVCs, have compact, $N({\HI})>10^{19}$~{\cmsq},
cold cores with few {\kms} FWHM surrounded by extended 
``halos'' with FWHM $\sim 25$~{\kms}.
The typical radius is $\sim5$--$8$~kpc at the estimated distance
of $\sim700$~kpc.  The BB sample is homogeneous but is not
complete; they estimate that there could be as many as
$200$ Local Group CHVCs.

Recently, Zwaan and Briggs (2000\nocite{zwaan}) reported
evidence in contradiction of the intragroup hypothesis.
In a blind {\HI} $21$--cm survey of extragalactic groups, sensitive to
$N({\HI}) \sim 10^{18}$~{\cmsq} (capable of detecting $\simeq
10^{7}$~M$_{\odot}$ {\HI} clouds), they failed to locate any
extragalactic counterparts of the Local Group HVCs.
This is in remarkable contrast to the numbers predicted.
If intragroup HVCs exist around all galaxies or galaxy groups, and
the {\HI} mass function is the same in extragalactic groups as measured
locally, then Zwaan \& Briggs should have detected $\sim70$ in groups
and $\sim 250$ around galaxies ($\sim 10$ and $\sim 40$ for the CHVCs,
respectively).

Thus, the Zwaan and Briggs\nocite{zwaan} result is in conflict with
the intragroup HVC hypothesis.
Since the hypothesized intragroup clouds are remnants of galaxy
formation and are shown to be stable against destruction mechanisms
(BSTHB\nocite{blitz}), they are predicted to form at very high redshifts 
and to be ubiquitous in galaxy groups to the present epoch.
In this {\it Letter}, we argue that the version of the intragroup HVC 
hypothesis presented by BSTHB is also in conflict with the observed
redshift number density of moderate redshift $(z \simeq 0.5)$ {\MgII} 
and Lyman limit (LLS) quasar absorption line systems.
We also find that the properties of the extragalactic analogs
of the BB CHVCs are severely constrained.

In general, the redshift number density of a non--evolving population
of objects, to be interpreted as the number per unit redshift, is written
\begin{equation}
\frac{dN}{dz} =  C_{f}\frac{n\sigma c}{H_0} 
\left( 1 + z \right) \left( 1 + 2q_{0}z \right) ^{-1/2} , 
\label{eq:dndz}
\end{equation}
where $C_{f}$ is the covering factor, $n$ is the number density of
absorbing structures, and $\sigma$ and $C_{f}$ are the surface area
presented by each structure and its covering factor for detectable
absorption.  Throughout, we use $H_{0} = 100$~{\kms}~Mpc$^{-1}$ and $q_0
= 0.5$, which gives $dN/dz \propto (1+z)^{1/2}$.

\section{M\lowercase{g} II Systems}
\label{sec:mgii}

The statistics of {\MgII} absorbers are well--established at $0.3 \leq
z \leq 2.2$.  
For rest--frame equivalent widths of $W({\MgII}) > 0.3$~{\AA}
(``strong'' {\MgII} absorption)
Steidel \& Sargent (1992\nocite{ss92}) found $dN/dz = 0.8\pm 0.2$ for
$z\simeq 0.5$ with a redshift dependence consistent with no evolution
expectations.
Normal, bright ($L \geq 0.1~L^{\ast}$) galaxies are almost
always found within $40$~kpc of strong {\MgII} absorbers
(\cite{bb91}; \cite{bergeron92}; \cite{lebrun93}; \cite{sdp94};
\cite{s95}; \cite{3c336}).
From the Steidel, Dickinson, and Persson survey, all but $3$ of $58$ strong {MgII} absorbers, 
detected toward $51$ quasars, have identified galaxies with a coincident
redshift within that impact parameter (sky projected separation
from the quasar line of sight) (see \cite{cc96}).
Also, it is rare to observe a galaxy with an impact parameter less than 
$\sim 40 h^{-1}$~kpc that does {\it not\/} give rise to {\MgII} 
absorption with $W({\MgII}) > 0.3$~{\AA} (\cite{s95}).  
In $25$  ``control fields'' of quasars, without observed strong {\MgII} 
absorption in their spectra, only two galaxies had impact parameters
less than $40 h^{-1}$~kpc (see also \cite{cc96}).
As such, the regions within $\sim 40 h^{-1}$~kpc of typical galaxies
account for the vast majority of {\MgII} absorbers above this
equivalent width threshold; there is nearly a ``one--to--one''
correspondence.
If we accept these results, it would imply that there is little
room for a contribution to $dN/dz$ from a population of intragroup
clouds {\it which have impact parameters much greater than $\sim40 h^{-1}$~kpc}.

However, the predicted cross section for {\MgII} absorption from 
the extragalactic intragroup clouds analogous to HVCs would be substantial.
We quantify the overprediction of the redshift path density by
computing the ratio of $dN/dz$ of the intragroup clouds to that of
{\MgII} absorbing galaxies,
\begin{equation}
\frac{(dN/dz)_{cl}}{(dN/dz)_{gal}} =  F 
  \left( \frac{f_{cl}}{f_{gal}} \right)
  \left( \frac{N_{cl}}{N_{gal}} \right) 
  \left( \frac{R_{cl}}{R_{gal}} \right) ^{2} ,
\label{eq:ratio}
\end{equation}
where $F$ is the fraction of {\MgII} absorbing galaxies that reside in
groups having intragroup HVC--like clouds, $f_{cl}$ is the fraction
of the area of the clouds and $f_{gal}$ is the fraction of the area
of the galaxies that would produce $W({\MgII}) > 0.3$~{\AA} along the 
line of sight, and $N_{cl}$ and $N_{gal}$ are the number of clouds 
and galaxies per group, respectively.
The cross section of the group times the intragroup cloud covering
factor, $C_{f}\cdot \pi R^{2}_{gr}$, is equal to $N_{cl} \cdot \pi R^{2}_{cl}$.
The total predicted $dN/dz$ for {\MgII} absorbers with $W({\MgII}) >
0.3$~{\AA} is then,
\begin{equation}
\left( \frac{dN}{dz} \right) _{tot} =
\left( \frac{dN}{dz} \right) _{gal}
\left[ 1 + \frac{(dN/dz)_{cl}}{(dN/dz)_{gal}} \right] .
\label{eq:totaldndz}
\end{equation}
If virtually all {\MgII} absorbers are accounted for by galaxies,
it is required that $(dN/dz)_{tot} \simeq (dN/dz)_{gal}$; the left
hand side of Equation~\ref{eq:ratio} must be very close to zero.

In the BSTHB version of the intragroup HVC model, the ``best''
expected values are $N_{cl}=300$ and $R_{cl} = 15$~kpc
(BSTHB; \cite{blitz-privcomm}); 
if we take $R_{gal} = 40$~kpc and $f_{gal} = 1$ (\cite{s95}),
and assuming $N_{gal} = 4$, we find that the covering factor for
{\MgII} absorption from extragalactic analogs to the Local Group HVCs
would exceed that from galaxies by a factor of $\sim 10$ for $F=1$ and
$f_{cl}=1$, giving $(dN/dz)_{tot} \simeq 9$.
More recently, Blitz and Robinshaw (2000\nocite{blitzdsph}) have
suggested that sizes may be smaller ($R_{cl} = 8$~kpc) when 
beam--smearing is considered.
Considering this as an indication of uncertainties in the
BSTHB parameters, and considering $2 < N_{gal} < 6$ for the
typical number of group galaxies, we find, for $F=1$ and
$f_{cl}=1$, a range $2 < (dN/dz)_{cl}/(dN/dz)_{gal} < 21$.
This corresponds to $2.6 < (dN/dz)_{tot} < 17.6$.
It is unlikely that $F$ is significantly less than unity;
the majority of galaxies reside in groups like the Local Group
that would have HVC analogs.
In order that $(dN/dz)_{tot} \sim (dN/dz)_{gal}$,
$f_{cl} \ll 0.2$ is required.

It is not clear what fraction $f_{cl}$ of
HVCs with $N({\HI})$ above the $10^{18}$~{\cmsq} detection 
threshold will give rise to $W({\MgII}) \geq 0.3$~{\AA}
because the equivalent width is sensitive
to the metallicity and internal velocity dispersion of the clouds.
Based upon Cloudy (\cite{ferland}) photoionization equilibrium models, 
a cloud with $N({\HI}) \simeq 10^{18}$~{\cmsq}, subject to
the ionizing metagalactic background (\cite{haardt-madau}), would give
rise to {\MgII} absorption with $N({\MgII}) \simeq
10^{14}N_{18}(Z/Z_{\odot})$~{\cmsq}, where $N_{18}$ is the {\HI}
column density in units of $10^{18}$~{\cmsq} and $Z/Z_{\odot}$ is the
metallicity in solar units.
For optically thick clouds, those with $N({\HI}) \geq
10^{17.5}$~{\cmsq}, this result is insensitive to the assumed
ionization parameter\footnote{The ionization parameter is the ratio of
the number density of hydrogen ionizing photons to the number density
of electrons, $n_{\gamma}/n_{e}$.} over the range $10^{-4.5}$ to
$10^{-1.5}$.

BSTHB expect HVCs to have $Z/Z_{\odot} \sim 0.1$.
For $N_{18} = 2$ and $Z/Z_{\odot} = 0.1$, clouds with internal
velocity dispersions of $\sigma _{cl} \geq 20$~{\kms} (Doppler
$b \geq 28$~{\kms}) give rise to $W({\MgII}) \geq 0.5$~{\AA}.
For $\sigma_{cl} = 10$~{\kms} ($b = 14$~{\kms}), $W({\MgII}) = 0.3$~{\AA}.
The CHVC ``halos'' typically have FWHM of $29$--$34$~{\kms}, which
corresponds to $\sigma _{cl} \sim 12$--$14$~{\kms} 
(\cite{braun-burton}).  
Thus it appears that most lines of sight through the BSTHB
extragalactic analogs will produce strong {\MgII} absorption.
Certainly $f_{cl} > 0.2$, so there is a serious
discrepancy between the predicted $(dN/dz)_{tot}$ and the
observed value.

However, if the intragroup clouds have lower metallicities, this would
result in smaller $W({\MgII})$.  Unfortunately, there has only been
one metallicity estimate published for an HVC, which may or may not be
related to the Galaxy.  Braun and Burton (2000\nocite{bb-hires}) estimate that
CHVC 125+41-207, with $W({\MgII}) = 0.15$~{\AA}, has a metallicity of $0.04
< Z/Z_{\odot} < 0.07$, however this is quite uncertain because of the 
effects of beam smearing on measuring the $N({\HI})$ value.
Because of the uncertainties, we simply state that a population of low
metallicity clouds could reduce the discrepancy
between the predicted redshift density for intragroup clouds,
$(dN/dz)_{cl}$, and the observed value of $(dN/dz)_{tot}$.  However,
then the expected number of smaller $W({\MgII})$ systems to arise from
intragroup clouds would be increased.

The observed {\MgII} equivalent width distribution rises rapidly below
$0.3$~{\AA} (``weak'' {\MgII} absorbers), such that $dN/dz =
2.2\pm0.3$ for $W({\MgII}) > 0.02$~{\AA} at $z=0.5$ (\cite{weak1}).
To this equivalent width limit, {\MgII} absorption could be observed
from intragroup HVCs with $N_{18}=2$ and metallicities as low as
$Z/Z_{\odot} = 0.0025$ [for $N({\MgII}) = 10^{11.7}$~{\cmsq},
$W({\MgII})$ is independent of $\sigma _{cl}$].  However, almost all
($9$ out of a sample of $10$) {\MgII} absorbers with $W({\MgII}) <
0.3$~{\AA} do {\it not\/} have associated Lyman limit breaks
(\cite{paper1}); that is, their $N({\HI})$ is more than a decade below
the sensitivity of $21$--cm surveys.  Thus, based upon available data,
roughly $90$\% of the ``weak'' {\MgII} absorbers do not have the
properties of HVCs, and therefore are {\it not\/} analogous to the
clouds invoked for the intragroup HVC scenario.  If $10$\% of the weak
{\MgII} absorbers are analogs to the intragroup HVCs, they would
contribute an additional $0.20$ to $(dN/dz)_{cl}$.

Since the BB CHVC extragalactic analogs have smaller cross sections,
we should separately consider whether they would produce a discrepancy
with the observed {\MgII} absorption statistics.  BB observed
$N_{cl}=65$ and inferred a typical $R_{cl} = 5$--$8$~kpc for the
CHVCs, however a complete sample might have $N_{cl}=200$.  Assuming
$N_{gal}=2$--$6$, $R_{gal} = 40$~kpc, $f_{gal} = 1$, and $F=1$, for
the BB subsample of the HVC population, we obtain 
$0.17 f_{cl} < (dN/dz)_{cl}/(dN/dz)_{gal} < 4.0 f_{cl}$.

The cores of the CHVCs have $N({\HI}) > 10^{19}$~{\cmsq} and they
occupy only about $15$\% of the detected extent.  
For $Z/Z_{\odot} > 0.01$ and $\sigma_{cl} = 10$~{\kms}, these cores can produce
$W({\MgII}) \ge 0.3$~{\AA} over their full area.  It follows that
$f_{cl} = 0.15$, which yields $0.025 < (dN/dz)_{cl}/(dN/dz)_{gal} <
0.6$.  Depending on the specific parameters, there may or may not be a
conflict with the observed $(dN/dz)_{tot}$ for strong {\MgII}
absorption.

The ``halos'' of the CHVCs have $N({\HI}) > 10^{18}$~{\cmsq} and, as
discussed above, would produce weak {\MgII} absorption for
$Z/Z_{\odot} > 0.005$ over most of the cloud area.  This implies
contribution to $(dN/dz)$ from BB CHVC analogs that is in the range
$0.14 < (dN/dz)_{cl} < 3.2$.  If the number were at the high end of
this range, the cross section would be comparable to the observed
$(dN/dz)$ for weak {\MgII} absorbers at $z=0.5$.  However, as noted
above when considering the BSTHB scenario, there is a serious
discrepancy.  Only $\sim 10$\% of the weak {\MgII} absorbers show a
Lyman limit break, so extragalactic analogs of the BB CHVCs can only
be a fraction of the weak {\MgII} population.  Regions of CHVCs at larger
radii, with $N({\HI})$ below the threshold of present $21$--cm
observations, are constrained to have $Z/Z_{\odot} \ll 0.01$ in order
that they do not produce a much larger population of weak {\MgII}
absorbers with Lyman limit break than is observed.

\section{Lyman Limit Systems}
\label{sec:lls}

The redshift number density of LLS also places strong constraints on
intragroup environments that give rise to Lyman breaks in quasar
spectra.  This argument is not sensitive to the assumed
cloud velocity dispersion and/or metallicity.

Statistically, $dN/dz$ for {\MgII} systems is consistent
(1~$\sigma$) with $dN/dz$ for LLS.
At $z \simeq 0.5$, LLS have $dN/dz = 0.5 \pm 0.3$ (\cite{kplls}) and
{\MgII} systems have $dN/dz = 0.8 \pm 0.2$ (\cite{ss92}).
Churchill \etal (2000a\nocite{paper1}) found a Lyman limit break
[i.e.\ $N({\HI}) \geq 10^{16.8}$~{\cmsq}] for each system in a sample
of ten having $W({\MgII}) > 0.3$~{\AA}.
LLS and {\MgII} absorbers have roughly the same redshift number
density and therefore {\MgII}--LLS absorption must almost always arise 
within $\sim 40$~kpc of galaxies (\cite{s93}).
As such, there is little latitude for a substantial contribution
to $dN/dz$ from intragroup Lyman limit clouds.

Using Equation~\ref{eq:dndz}, we could estimate this contribution by
considering the volume density of galaxy groups and the cross section
for HVC Lyman limit absorption in each.
However, the volume density of groups is not well measured,
particularly out to $z=0.5$.

Instead, we make a restrictive argument based upon a comparison
between the cross sections for Lyman limit absorption of $L^{\ast}$ 
galaxies and for HVCs in a typical group (similar to the discussion of
{\MgII} absorbers in \S~\ref{sec:mgii}).
Again, we simply compare the values of $C_f$ for the different
populations of objects in a typical group.
The covering factor for HVCs within the group is
\begin{equation}
C_f = N_{cl}
\frac{\left(R_{cl}\right)^2}{\left(R_{gr}\right)^2} .
\end{equation}
The best estimate for the BSTHB version of the intragroup HVC model,
with $N_{cl} = 300$, $R_{cl} = 15$~kpc, and a group radius 
$R_{gr} = 1.5$~Mpc, gives $C_{f} = 0.03$ for $N({\HI}) \simeq 2 \times
10^{18}$~{\cmsq}.  
If instead we use the BB number of observed CHVCs, $N_{cl} = 65$,
and $R_{cl} = 5$--$8$~kpc, we obtain a much smaller number, 
$0.0007 < C_{f} < 0.0018$.
However, if the BB sample is corrected for incompleteness such that
$N_{cl} = 200$, these numbers increase so that
$0.002 < C_{f} < 0.006$.

In comparison, a typical group with $\sim4$ $L^{\ast}$ galaxies, each with a
Lyman limit absorption cross section of $R_{cl} \sim 40$~kpc, would
have $C_f = 0.002$.  
If they existed with the properties discussed,
the extragalactic analogs to the BSTHB HVCs would dominate the
contribution of $L^{\ast}$ galaxies to the $dN/dz$ of LLS by at least
a factor of $\sim 15$, and this is only considering HVC regions with
$N({\HI}) > 2 \times 10^{18}$~{\cmsq} that are detected in the $21$--cm
surveys.  
Any extensions in area below this threshold value [down to
$N({\HI}) \sim 5 \times 10^{16}$~{\cmsq}] would worsen the discrepancy.
As such, the BSTHB hypothesis is definitively ruled out.

For regions of BB CHVCs with $N({\HI}) > 2 \times 10^{18}$~{\cmsq}, the
covering factor ranges from $C_f =0.0007$ to $C_f=0.006$ depending
on assumed sizes and corrections for incompleteness.  This ranges
from $35$--$300$\% of the cross section for the $L^{\ast}$ galaxies.
The {\it total} observed $dN/dz$ for Lyman limit absorption 
(down to $\log N({\HI}) = 17$~{\cmsq}) is only $\sim 0.5$, even a $35$\% 
contribution to the Lyman limit cross section from HVCs 
that are separate from galaxies creates a discrepancy.
This would imply that the result that most lines of sight within 
$40$~kpc of a typical $L^{\ast}$ galaxy produce Lyman limit
absorption is incorrect.
This would further imply that there is a substantial population of
strong {\MgII} absorbers without Lyman limit breaks
(to account for $dN/dz = 0.8$ for strong {\MgII}
absorption) or of strong {\MgII} absorbers not associated
with galaxies.  
Both types of objects are rarely observed
(\cite{paper1}; \cite{bb91}; \cite{bergeron92}; \cite{lebrun93};
\cite{sdp94}; \cite{s95}; \cite{3c336}).
Furthermore, $C_f=0.002$ for BB CHVCs only takes into account the
fraction of the BB CHVC areas with $N({\HI}) > 2 \times 10^{18}$~{\cmsq}.
Therefore, the extended ``halos'' around the CHVCs are also
constrained not to contribute substantial cross section for Lyman
limit absorption along extragalactic lines of sight.

\section{Summary}
\label{sec:summary}

We have made straight--forward estimates of the
predicted redshift number density at $z \simeq 0.5$ of
{\MgII} and LLS absorption from hypothetical extragalactic
analogs to intragroup HVCs as expected by extrapolating from the BSTHB
and BB Local Group samples.  We find that it is difficult to reconcile
the intragroup hypothesis for HVCs with the observed $dN/dz$ of
{\MgII} and LLS systems.

The discrepancy between the $dN/dz$ of {\MgII}--LLS absorbers and the
observed covering factor of ``intragroup'' HVCs could be reduced if
the HVCs have a clumpy structure.  Such structure would result in
{\MgII}--LLS absorption observable only in some fraction, $f_{los}$,
of the lines of sight through the cloud.  Effectively, this reduces
the covering factor for Lyman limit absorption, or the value of
$f_{cl}$ in equation (2) for {\MgII} absorbers.  Considering beam smearing
in $21$--cm surveys, substructures would be detected above a
$N({\HI}) > 2 \times 10^{18}$~{\cmsq} $21$--cm detection threshold
if their column densities were
$N({\HI})_{sub} > 2 \times 10^{18}/f_{los}$~{\cmsq}.  The predicted
$dN/dz$ for HVC--like clouds could be reduced by a factor of ten if
$f_{los} \leq 0.1$, giving $N({\HI})_{sub} > 2 \times
10^{19}$~{\cmsq}.  All the gas outside these higher {\HI} column
density substructures would need to be below the Lyman limit or the
arguments in \S~\ref{sec:lls} would hold.  It is difficult to
reconcile such a density distribution with the high resolution
observations of BB CHVCs which show diffuse halos around the
core concentrations (\cite{bb-hires}), but these ideas merit
further consideration.

\subsection{The BSTHB Scenario}

We conclude that the predicted $dN/dz$ from the hypothetical
population of intragroup HVCs along extragalactic sight lines to
quasars from the BSTHB scenario would exceed: 

1) the $dN/dz$ of {\MgII} absorbers with $W({\MgII}) \geq 0.3$~{\AA}.
This class of absorber is already known to arise within $\sim 40 h^{-1}$~kpc
of normal, bright galaxies (\cite{bb91}; \cite{bergeron92};
\cite{lebrun93}; \cite{sdp94}; \cite{s95}; \cite{3c336}).

2) the $dN/dz$ of ``weak'' {\MgII} absorbers with $0.02 < W({\MgII}) <
0.3$~{\AA} absorption.  
In principle, weak {\MgII} absorption could
arise from low metallicity, $0.005 \leq Z/Z_{\odot} < 0.1$, intragroup
HVCs.  However, the majority of observed weak systems are already
known to be higher metallicity, $Z/Z_{\odot} \simeq 0.1$, sub--Lyman
limit systems (\cite{weak1}; \cite{rigby}).

3) the $dN/dz$ of Lyman limit systems.
These would be produced by all extragalactic BSTHB HVC analogs
regardless of metallicity.
However, {\it most\/} Lyman limit systems are seen to arise within 
$\simeq 40 h^{-1}$~kpc of luminous galaxies (\cite{s93}; \cite{paper2}).

These points do not preclude a population of infalling intragroup
clouds which do not present a significant cross section for
$21$--cm absorption, as predicted by CDM models
(\cite{iforget}; \cite{moore}).
In fact, such intragroup objects could be related to sub--Lyman limit
weak {\MgII} absorbers (\cite{rigby}).

\subsection{The BB Scenario}

The properties of the BB CHVC population are also significantly
constrained by {\MgII} and Lyman limit absorber statistics:

1) They {\it could\/} produce $W({\MgII}) \geq 0.3$~{\AA} in excess of
what is observed if a large incompleteness correction is applied
(i.e. so that $N_{cl} = 200$), or if relatively large sizes ($R_{cl}
\sim 8$~kpc) are assumed.

2) They would be expected to contribute to the $dN/dz$ of
weak [$W({\MgII}) > 0.02$~{\AA}] {\MgII} absorption.
However, based upon observations (\cite{paper1}), only $\sim 10$\% 
of the population of weak {\MgII} absorbers have Lyman--limit breaks.
Therefore, only a small fraction of weak {\MgII} absorption could 
arise in extragalactic BB CHVC analogs.

3) The $dN/dz$ for Lyman limit absorption from the hypothesized BB
CHVC population could be a significant
fraction, or comparable to that expected from the local environments of
$L^{\ast}$ galaxies (within $40$~kpc); the observed value is already
consistent with that produced by the galaxies.

4) The CHVCs are observed to have a cool core with $N({\HI}) >
10^{19}$~{\cmsq}, surrounded by a halo which typically extends to
$R_{cl} \sim 5$~kpc.  It is natural to expect that the {\HI} extends
out to larger radii at smaller $N({\HI})$ and should produce a Lyman
limit break out to the radius at which $N({\HI}) < 10^{16.8}$~{\cmsq}.
Although there is expected to be a sharp edge to the {\HI} disk at
$N({\HI}) \sim 10^{17.5}$ or $10^{18}$~{\cmsq} (\cite{maloney};
\cite{corbelli}; \cite{dove}), physically we would expect that this
edge would level off at $\sim 10^{17.5}$~{\cmsq}, such that a
significant cross section would be presented at $10^{16.8} < N({\HI})
< 10^{17.5}$~{\cmsq}.  Another possibility is that there is a sharp
cutoff of the structure at $N({\HI}) \sim 2 \times 10^{18}$~{\cmsq},
but this is contrived.

\section{Conclusion}

We are forced to the conclusion that there can only be a limited
number of extragalactic infalling group HVC analogs at $z \sim 0.5$.
Future data could force a reevaluation of the relationships between
galaxies, Lyman limit systems, and {\MgII} absorbers, but it seems
unlikely that the more serious inconsistencies we have identified
could be reconciled in this way.  A clumpy distribution of {\HI}
could be constructed that would reduce the discrepancy, but would
require very diffuse material (below the Lyman limit) around dense 
cores.  Evolution in the population of HVCs
is another possibility.  If the extragalactic background radiation
declined from $z=0.5$ to the present, the clouds would have been more
ionized in the past, and therefore would have had smaller cross
sections at a given $N({\HI})$.  However, this does not explain why
Zwaan and Briggs (2000\nocite{zwaan}) do not see the $z=0$
extragalactic analogs to the HVCS or CHVCs.  Our results are entirely
consistent with theirs, and the implications are the same: the
discrepancies between the Local Group HVC population and the
statistics of {\MgII} and Lyman limit absorbers can only be reconciled
if most of the extragalactic HVC analogs are within $100$--$200$~kpc
of galaxies, and not at large throughout the groups.

\acknowledgements
We thank L. Blitz, J. Bregman, J. Mulchaey, B. Savage, K. Sembach,
T. Tripp, and especially B. Wakker and Buell Jannuzi, and our
referees for stimulating discussions and comments.
Support for this work was provided by NSF grant AST--9617185
(J. R. R. was supported by an REU supplement) and by NASA grant NAG
5--6399.  


\end{document}